\documentclass[12pt,twoside]{article}   
\usepackage{latexsym} 
\usepackage{mathptm}
\usepackage{amsmath}
\usepackage{amssymb}
\usepackage{graphicx}

\begin{document}
 
\title{Comment on ``No-go theorem for bimetric gravity with positive and negative mass''}
\author{Sabine Hossenfelder \thanks{sabineh@nordita.org}\\
{\footnotesize{\sl Nordita, Roslagstullsbacken 23, 106 91 Stockholm, Sweden}}}
\date{}
\maketitle
 
\vspace*{-1cm}
\begin{abstract}
Authors Hohmann and Wohlfarth have put forward a no-go theorem for bimetric gravity with positive and negative 
mass in arXiv:0908.3384v1 [gr-qc]. This comment shows that their no-go theorem does not apply to arXiv:0807.2838v1 [gr-qc].

\end{abstract}

In a recent paper \cite{Hohmann:2009bi} authors Hohmann and Wohlfarth 
present a no-go theorem for bimetric gravity with positive and negative mass. Based on five assumptions 
they proof that in the Newtonian limit a bimetric theory fulfilling these assumptions cannot lead to
a Poisson equation that weights standard matter in the usual way and a conjectured new `anti-gravitating' type
of matter with the opposite sign though the same coupling constant. 

A bi-metric model with exchange symmetry that allows for both positive and negative gravitational masses
has been discussed in \cite{Hossenfelder:2008bg}. It should be emphasized
that the inertial masses in the model presented in \cite{Hossenfelder:2008bg} remain positive. This is also the case considered
in \cite{Hohmann:2009bi}. 

In \cite{Hohmann:2009bi}, the both metrics are denoted $g^+$ and $g^-$, which corresponds to the tensors ${\bf g}$ and
${\bf \underline h}$ in \cite{Hossenfelder:2008bg}. The anti-gravitating matter is introduced in both papers 
as an entirely new matter sector that interacts only gravitationally with our normal matter (and is thus dark). Since gravity is
mediated by a spin-2 field, like charges attract and unlike charges repel. The different motion of the
anti-gravitating matter is obtained by a second covariant derivative that preserves the second metric. 
With respect to our usual metric, this connection is torsion-free but non-metric. 

The introduction of the
second connection suffices to describe the motion of anti-gravitating test-particles in a 
background field, but to determine
the effect of matter sources on the geometry the field equation have to be found. These field equations
must have additional source terms stemming from the new kind of matter. There is however no reason why
the additional anti-gravitating sources for the metric ${\bf g}$ should be conserved with respect to the
ordinary derivative, since, as a consequence of a symmetry argument, the equations of motion for the anti-gravitating matter contain only the second
derivative which is non-metric with respect to ${\bf g}$. Without further attention, this would lead to an 
inconsistency since the Bianchi-Identities for the curvature were then not complemented by a conservation of the source. 

Another way to see this is that the field equations are 10 equations for 10 components of the metric,
related by the four contracted Bianchi-Identities, leaving 4 degrees of freedom for the choice of
the coordinate system. If the source has an additional term that it not automatically conserved by
virtue of conservation of energy, then the Bianchi-Identities add 4 additional constraints. These cannot
in general be fulfilled since we have degrees of freedom missing. This has been discussed in 
detail in \cite{Hossenfelder:2008bg}, but is missing in the ``naive counting argument'' of assumption (i) in \cite{Hohmann:2009bi}. 

The solution to this dilemma is to realize that observers of both types of matter 
must be able to choose their
coordinate systems independently, and there is no way for them to compare their
measuring sticks (no device to contract one local basis with the other). Thus,  
there are additional degrees of freedom in mapping tensors describing observables of the
one type of matter to that of the other and vice versa. This leads to the introduction
of two functions, called the `pull-overs' in \cite{Hossenfelder:2008bg} (denoted $P_{\underline h}$ and $P_g$), that each 
carry 4 degrees of freedom (after requiring them to respect tensorial structure and the torsion-free-ness of the connections). 
The equations of motion in  \cite{Hossenfelder:2008bg} thus are not of the form of assumption (i) in the proof of \cite{Hohmann:2009bi}, and it is in fact doubtful the assumption (i) in general constitutes a self-consistent set of equations (it certainly does in special cases).

More important than this is however assumption (iii) in \cite{Hohmann:2009bi}. It states that the source terms
do not mix both types of matter, reflected in the matrix ${\underline{\underline J}}$ being diagonal. 
In addition the coefficients in that matrix are assumed to be constant. In \cite{Hossenfelder:2008bg}, the coefficients
of that matrix are not constant, as one can read off Eqs (34) and (35) in \cite{Hossenfelder:2008bg}.  One reason for this is that the sources are densities, and if they are defined
through variation with respect to the corresponding metric, the measure has to be converted when the source couples
to the other metric. In addition, the ${\underline{\underline J}}$ that would be extracted from that model also contains the pull-overs that have been neglected in \cite{Hohmann:2009bi} altogether. However, in the Newtonian limit, these coefficients are constant indeed and since the Newtonial limit is what the authors of \cite{Hohmann:2009bi}
 are eventually interested in, assuming constancy in general is careless but not the relevant point. 

The relevant point is their assumption that the matrix ${\underline{\underline J}}$ be diagonal,
which is ``easily motivated by recalling that the [usually gravitating fields] should only couple
to the [usual metric], and the [anti-graviating] fields only to the [second metric].'' With this
motivation, the assumption follows from the action. The action of the model presented in \cite{Hossenfelder:2008bg}  is
not of this form, see Eq (32) in \cite{Hossenfelder:2008bg}; the second term couples the anti-gravitating field to
the usual metric, while the fourth term couples the normally gravitating field to the second metric (a consequence
of the pull-overs in these terms). 

Indeed, if one takes into account that theories with more than one interacting spin-2 field are known to be 
inconsistent already \cite{Boulanger:2000rq}, the graviational part of the action should not couple both 
metrics, and thus the graviational
tensor $\underline K$ in the field equations should not mix the both metrics either. It would then not be too surprising 
that the assumption the anti-gravitating matter does not couple to the usual metric does eventually lead to a contradiction with a Newtonian limit in which the anti-gravitating source enters the Newtonian potential that appears
to first order in the metric. It is interesting however that \cite{Hohmann:2009bi} arrives at this conclusion
without further restriction on the interaction of the two metrics.

However, in the Newtonian limit of the model in \cite{Hossenfelder:2008bg}, the matrix ${\underline{\underline J}}$ is not invertible
as can easily be read off Eqs (34) and (35) in \cite{Hossenfelder:2008bg}. Instead, in the Newtonian limit, 
$V, {\underline W}, g, {\underline h} \to 1$ and the $a^{\nu}_{\;\;\underline \kappa}$ and their inverse
are just the identity, meaning  ${\underline {\underline J}}$  has entries $1,-1$ in both the first and second row.
Thus, the crucial conclusion following Eq. (31) in \cite{Hohmann:2009bi}
cannot be made and the proof fails.

\section*{Acknowledgements}

I want to thank Manuel Hohmann and Mattias Wohlfarth for their interest, and Manuel
Hohmann for an interesting and clarifying exchange. 
\end{document}